\documentclass[sigconf]{acmart}

\renewcommand\footnotetextcopyrightpermission[1]{}
\AtBeginDocument{
  }

\setcopyright{acmlicensed}
\copyrightyear{2018}
\acmYear{2018}
\acmDOI{XXXXXXX.XXXXXXX}

\acmConference[Conference acronym 'XX]{Make sure to enter the correct
  conference title from your rights confirmation emai}{June 03--05,
  2018}{Woodstock, NY}
\acmISBN{978-1-4503-XXXX-X/18/06}

\usepackage{graphicx}
\usepackage{listings}
\usepackage{natbib}
\usepackage{multirow}
\usepackage{amsmath}
\usepackage{fancyhdr}
\usepackage{lineno}
\usepackage{url}

\newtheorem{theorem}{Theorem}
\newtheorem{corollary}{Corollary}

\usepackage{comment}

\begin{document}

\title{An Actor-Critic Approach to Boosting Text-to-SQL Large Language Model}

\author{Ziyang Zheng}
\email{107552201369@stu.xju.edu.cn}
\affiliation{
  \institution{Xinjiang University}
  \city{Urumqi}
  \state{Xinjiang}
  \country{China}
}

\author{Haipeng Jing}
\email{107552304072@stu.xju.edu.cn}
\affiliation{
  \institution{Xinjiang University}
  \city{Urumqi}
  \state{Xinjiang}
  \country{China}
  }

\author{Canyu Rui}
\email{20221401240@stu.xju.edu.cn}
\affiliation{
  \institution{Xinjiang University}
  \city{Urumqi}
  \state{Xinjiang}
  \country{China}
  }

\author{Askar Hamdulla}
\email{askar@xju.edu.cn}
\affiliation{
  \institution{Xinjiang University}
  \city{Urumqi}
  \state{Xinjiang}
  \country{China}
  }

\author{Dong Wang}
\email{wangdong99@mails.tsinghua.edu.cn}
\affiliation{%
  \institution{Tsinghua University}
  \city{Haidian Qu}
  \state{Beijing Shi}
  \country{China}
  }

%% \renewcommand{\shortauthors}{Trovato et al.}

\begin{abstract}
  Text-To-SQL (T2S) conversion based on large language models (LLMs) has found a wide range of applications, by leveraging the capabilities of LLMs in interpreting the query intent expressed in natural language. Existing research focuses on suitable representations for data schema and/or questions, task-specific instructions and representative examples, and complicated inference pipelines. All these methods are empirical and task specific, without a theoretical bound on performance. In this paper, we propose a simple, general, and performance guaranteed T2S enhancement approach called Actor-Critic (AC). Specifically, we design two roles using the same LLM: an Actor to produce SQL queries and a Critic to evaluate the produced SQL. If the Critic believes the produced SQL is wrong, it notifies the Actor to reproduce the SQL and perform evaluation again. By this simple iterative process, expected performance can be derived in theory. We conducted extensive experiments on the Spider and related datasets with eleven LLMs, and demonstrated that the Actor-Critic method consistently improves the performance of T2S, thus serving as a general enhancement approach for T2S conversion.
\end{abstract}




\maketitle

\section{Introduction}

Text-to-SQL (T2S) is a significant research direction in the field of natural language processing (NLP), with the aim of transferring natural language queries into structured query language (SQL) statements, thereby enabling non-expert users to interact with databases using natural language~\cite{deng2022recent,katsogiannis2023survey}. Most previous work has focused on learning a task-specific transfer model that converts questions in natural language to SQL statements with supervised learning, usually by fine-tuning a pre-trained model to tackle the problem of data scarcity~\cite{bogin2019representing,wang2020rat,li2023resdsql}. 

In recent years, the emergence of decoder-based large language models (LLMs) has established a unified solution for various NLP tasks~\cite{radford2019language,brown2020language,touvron2023llama}. By designing appropriate prompts, LLMs can solve multiple tasks in a zero-shot manner~\cite{radford2019language}, and performance can be further improved with a few examples to demonstrate the task~\cite{brown2020language}, in particular with examples to demonstrate the inference process, aka. the chain of thought (CoT)~\cite{wei2022chain}. More complex problems can be solved by prompting LLMs in a hierarchical way, for instance with the least-to-most approach~\cite{zhouleast} that lets the LLM decompose the task into simpler subtasks and then solves the subtasks step by step following the LLM's instruction. 

The potential of LLMs in zero-shot learning and few-shot learning has been utilized in the text-to-SQL task~\cite{liu2023comprehensive,rajkumar2022evaluating,gao2024text,zhang2024benchmarking}. A large body of research in this direction focuses on how to prompt LLMs using rich-information prompts. This is 'prompt engineering' includes powerful representations for database schema and questions in the zero-shot scenario~\cite{chang2023prompt,dong2023c3,pourreza2024din} and task-dependent example selection in the few-shot scenario~\cite{guo2023case,liu2022makes,nan2023enhancing}. 
Recently, inspired by the idea of CoT and Least-to-Most, some research focuses on constructing complex inference pipelines by demonstration, in order to solve complex tasks requiring nested or hierarchical solving processes. For instance, DIN-SQL~\cite{pourreza2024din} instructs LLMs to decompose complex queries into simpler query subtasks, and then solve the subtasks sequentially. Note that this pipeline construction requires a demonstration of the inference process, thus belonging to the category of few-shot learning. 
By carefully designing the prompts, high text-to-SQL performance has been reported on both open-source models such as Dolly~\cite{DatabricksBlog2023DollyV2,sun2023battle}, 
LLaMA~\cite{touvron2023llama,touvron2023llama2,llama32024,sun2023battle,gao2024text}, 
Vicuna~\cite{vicuna2023,sun2023battle,gao2024text} and Guanaco~\cite{dettmers2024qlora,sun2023battle}, 
as well as commercial models such as GPT-3.5~\cite{gpt35,hong2024knowledge} and GPT-4o~\cite{gpt4o,hong2024knowledge}.

%wang2023mac: self correction, sq execution
%liu2023comprehensive: Liu A, Hu X, Wen L, et al. A comprehensive evaluation of ChatGPT's zero-shot Text-to-SQL capability[J]. arXiv preprint arXiv:2303.13547, 2023. [zero short]
%li2024pet： PET-SQL: A Prompt-enhanced Two-stage Text-to-SQL Framework with Cross-consistency
%rajkumar2022evaluating： Evaluating the Text-to-SQL Capabilities of Large Language Models [Codex model,zero shot and few shot]
%sun2023battle: open-source models are still bad
\begin{figure*}
    \centering
    \includegraphics[width=\linewidth]{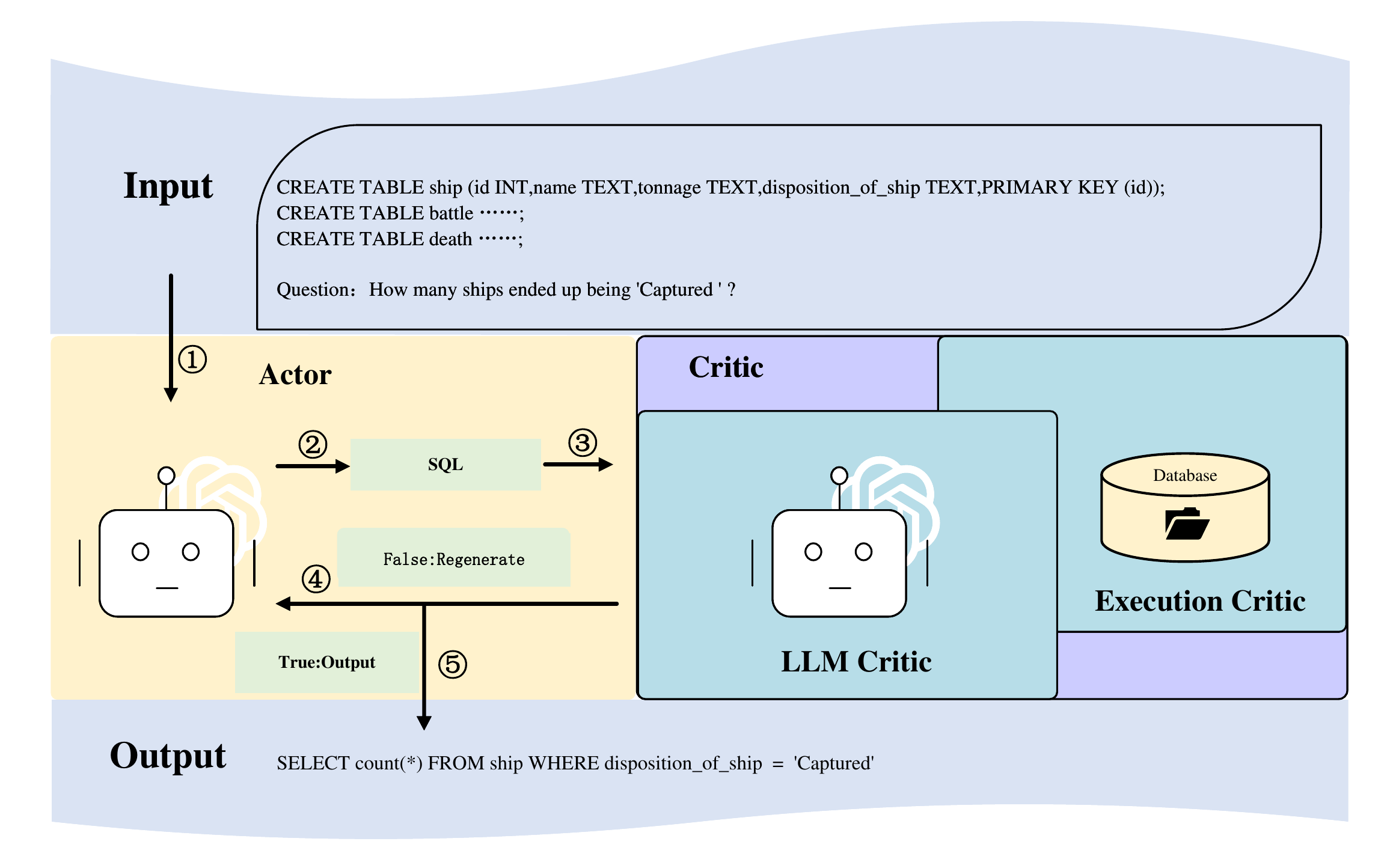}    \caption{The framework of AC-SQL. (1) The input involves the SQL statements for schema creation and the question in natural language. (2) The Actor takes this input and generates an SQL candidate. (3) The SQL candidate, together with the input to the Actor, is forwarded to the Critic, where the SQL candidate is evaluated by consulting an LLM (LLM Critic) and accessing the database (Execution Critic). (4) If the output of the Critic is False, the Actor is signified to regenerate the SQL candidate; (5) If the output of the Critic is True, the SQL candidate is output. }
    \Description{The framework of AC-SQL.}
    \label{fig:arch}
\end{figure*}

In spite of the promising results, most of the present prompt engineering methods are tightly correlated with the model and/or the query task, leading to poor generalization. For instance, the prompts designed to instruct an LLM  to generate syntactically valid SQL statements may not work with another LLM, and the demonstration examples, no matter the question-SQL pairs or the inference pipeline, are task-related. 
This means that the performance with the deliberately designed prompting methods is unpredictable when employed to real applications, and their effectiveness needs to be carefully experimented with. In real applications, we prefer a general approach that can deliver performance improvement safely without knowing the details of the model and the task. 

In this paper, we present an \textbf{Actor-Critic (AC) approach} that can safely improve the performance of LLM-based text-to-SQL systems without any knowledge of the model and the query task. Specifically, we set up two agents, one is called Actor and the other is called Critic. These two agents can solve the text-to-SQL task in both a collaborative and competitive way, as shown in Fig.~\ref{fig:arch}. More specifically, the Actor produces a SQL candidate, while the Critic evaluates the candidate and verifies if it is correct in both syntactic form (by an Execution Critic) and semantic meaning (by an LLM Critic). If the Critic returns True, the SQL statement will be returned as the output; otherwise the Actor will be signified to regenerate a new candidate. This process iterates until the Critic outputs True or a pre-defined number of iterations is reached. 

The key idea is the relationship between problem solving and verification. In the theory of computational complexity, the problems whose solutions can be \emph{effectively} produced and verified by methods form the P set and NP set, respectively~\cite{karp2010reducibility}. Our assumption is that solving and verification are different tasks and require different intelligent power, and so the accuracy of the solution can be improved by a solving-and-verifying iteration, which we call the `solving-and-verifying assumption'.
It should be noted that 'solving and verification' is a common practice in our daily life. For instance, in a math examination, students usually find a solution first, and then recheck the solution to make sure it is correct. 
The AC-SQL approach simulates this solving-and-verifying mental process, by integrating the capability of LLMs in SQL generation and verification. We highlight that this AC-SQL approach is \emph{generic}, meaning that it can be applied to any task and any model; moreover, it can be shown that its performance is \emph{guaranteed}, i.e., its expected performance can be derived in theory.

We evaluated the proposed AC approach for text-to-SQL, named AC-SQL from now on, on the widely used Spider dataset and its two variants. To evaluate the generality of the approach, our experiments were designed following three principles: (1) We focused on the zero-shot paradigm, i.e., without any demonstration samples as in the few-shot paradigm. The first reason is that demonstration samples are not always available in real applications, and the second reason is that the few-shot demonstrations and the AC-SQL approach are orthogonal techniques. (2) We focused on simple prompts, as complex prompts tend to be more dependent on tasks and models. Moreover, with the increasing of the power of LLMs, complex prompts will be less required.  (3) We focused on simple pipelines, preventing complex guidance for the inference procedure. Our experiments showed that the AC-SQL method resulted in a consistent and very significant performance improvement in execution accuracy with a broad spectrum of open-source models, and it is also an effective approach with the most powerful commercial models so far, including GPT-3.5 and GPT-4o.

The contribution of the paper is summarized as follows: (1) We present a simple, generic, and performance guaranteed AC-SQL approach to boost LLM-based text-to-SQL performance. (2) We conducted a comprehensive study on the proposed AC-SQL approach, with a broad range of LLMs and query tasks, and demonstrated that it can lead to consistent and significant gains even with the most powerful GPT-4o model.

\section{Related Work}

\subsection{Text-to-SQL research}

Previous Text-to-SQL methods primarily relied on supervised learning, which encodes information of the schema and the question by an encoder, and produces SQL statements by a decoder. This approach requires a large amount of training data, i.e., question-SQL pairs, that is not practical in most applications~\cite{guo2019towards,choi2021ryansql,wang2020rat}.

In contrast, the newly emerged LLM-based methods require no or very few text-SQL pairs, thus highly promising for real applications. The first strong LLM for Text-to-SQL is Codex~\cite{chen2021evaluating}, which was specifically tuned to generate codes. Empirical studies showed that Codex can achieve fairly good performance on the Spider benchmark, even approaching to results obtained by supervised learning~\cite{rajkumar2022evaluating}. The advantage on code generation has been inherited by GPT-3.5 and GPT-4~\cite{dong2023c3,liu2023comprehensive}, as demonstrated by the clear advantage of GPT models on Text-to-SQL tasks compared to open-source models~\cite{sun2023battle}.

Recently, a large body of research followed the LLM-based method, roughly categorized into three research lines. The first research line is to seek more suitable representations for the data schema and question~\cite{chang2023prompt,dong2023c3,pourreza2024din}, to assist the LLMs in understanding the structure of the database and the intention of the question. The second research line is to select suitable examples to demonstrate the task~\cite{gao2024text,guo2023case,liu2022makes,nan2023enhancing}. For instance, DAIL-SQL~\cite{gao2024text} selects examples by looking at the similarity between the input question and the questions in the example set. The third direction is to design complex pipelines to tackle questions requiring complicated inference.  
For instance, in DIN-SQL~\cite{pourreza2024din}, the question is first analyzed to identify relevant schema, and then the question is classified into three different categories with different complexities, by which different prompting methods are used. In particular, for complex SQL with nested queries, CoT was employed by few-shot demonstration. This design leverages the capacity of LLMs in finding relevant schema and classifying questions, and alleviates the potential harm of complex reasoning methods such as CoT with simple questions. MAC-SQL~\cite{wang2023mac} implements a similar approach. 

%In C3 zero-shot prompting method~\cite{dong2023c3}, They also introduced the concept of self-consistency, greatly reducing the impact of LLM instability~\cite{dong2023c3}. 
%Researchers also compared the parsing abilities of six popular LLMs, including LLaMA, Vicuna, and ChatGPT, on nine benchmark datasets, providing valuable insights for researchers in selecting models for Text-to-SQL technology (Sun et al., 2023) \cite{sun2023battle}.

The AC-SQL approach presented in this paper is orthogonal to existing methods, and can be employed with existing methods together to gain additional performance improvement.

\subsection{Boosting by Verification}

We also note that the idea of verifying the generated SQL statements has been investigated in several previous studies. For instance, DIN-SQL~\cite{pourreza2024din} presented a self-correction approach that asks the LLM to refine the SQL. This is similar to the LLM-Critic in AC-SQL, but the difference is that the LLM-Critic in AC-SQL simply focuses on verifying the solution, leaving SQL generation as the professional task of the Actor. This design tends to be more stable as both Actor and Critic focus on their own distinct tasks, and more importantly, a theoretical performance guarantee can be derived. 
\cite{sui2023reboost} presented a verification stage by contacting the database API. Similarly, in the C3 approach~\cite{dong2023c3}, multiple SQL candidates were generated and executed via the database API, and the SQL statements that yield the most consistent output were regarded to be correct. All these methods are similar to the Execution-Critic in AC-SQL, though there is no interaction between the generator and the critic. 

\subsection{Critics for LLMs}

Critics have been widely used in training human-aligned LLMs. Perhaps the most representative example is reinforcement learning from human feedback (RLHF) 
\cite{christiano2017deep}, where human preference was used as critics to train a
reward function that was then used to approximate human's judgement to align a pre-trained LLM~\cite{ouyang2022training,openai2024gpt4technicalreport}. A central idea behind RLHF is that evaluation of solutions are often faster and easier for humans than demonstration of solutions~\cite{leike2018scalable}. The same idea was employed to train more advanced critics, i.e., critics that not only verify correctness or provide preference, but also describe flaws in solutions. In this setting, human provides feedback about the correctness of the critics, forming a two-level hierarchy of critics, i.e., critics' critics~\cite{saunders2022self,mcaleese2024llmcriticshelpcatch}. Theoretical analysis was conducted based on the complexity theory~\cite{irving2018ai,saunders2022self}, indicating that models trained with n-step critics (including human at the final step) can solve problems analogous to PSPACE~\cite{shamir1992ip}. Despite the inspiration of these theoretical work, it should be noted that the assumption behind the analysis is human can provide correct feedback (analogous to a polynomial time Turing machines) and the model can be perfectly trained so that the solutions/critics do not encountered any critical opinion by the upper-level critic. If this is not the case, the solution is not guaranteed to be correct. 

Another approach that utilizes critics to align LLMs to human judgement is the popular best-of-N sampling~\cite{stiennon2020learning,nakano2021webgpt,jinnai2024regularized}, which produces $N$ solutions from an LLM and selects the best solution according to a critic. An advantage of this approach is it does not require to fine-tune the base model for the solution, and the base model can be a black-box. A shortcoming is that the critic needs to produce a criticizing score to represent the correctness of a solution, which is not easy if the critic is an LLM. 

AC-SQL can be regarded as a 1-step critic hierarchy, thus it can solve NP problems if the Critic is perfect. It is different from the alignment training methods as it does not fine-tune the base model, but works at inference time. It is also different from the best-of-N sampling as it is based on a True/False Critic implemented using an LLM. Essentially, AC-SQL boosts performance of SQL statements generation by leveraging the prior probability that the Critic yields correct judgement.

%\cite{mcaleese2024llmcriticshelpcatch} Train a critic to critic code writing with RLHF
%\cite{saunders2022self} It fine-tune large language models to write natural language critiques (natural language critical comments) using behavioral cloning. Larger models can also integrate their own selfcritiques as feedback, refining their own summaries into better ones.

%\cite{amodei2016concrete} 

\section{Method}

\subsection{Task definition}

We follow the formulation in~\cite{liu2023comprehensive,sun2023sql} to formalize the LLM-based text-to-SQL task. Given a question $Q$ in natural language, a database schema $D$, a prompt $P$ is designed to instruct an LLM $M$. The goal of the Text-to-SQL task is to maximize the probability of the correct SQL statement $Y$ corresponding to the question $Q$. Suppose $M$ yields $Y$ token by token, the probability can be written as follows:

\begin{equation}
 \log P_M(Y|Q,D,P) = \sum_{i=1}^{|Y|} \log P_M(Y_i|Q,D,P,Y<i)
\end{equation}

\noindent where $Y<i$ represents the prefix of the SQL $Y$ up to the position $i-1$; $P_M(Y_i|Q,D,P,Y<i)$ is the conditional probability of the $i$-th token in SQL query $Y$ given the prefix $Y<i$, prompt $P$, database $D$, and question $Q$.

\subsection{Implementation of AC-SQL}

The AC-SQL approach is shown in Fig.~\ref{fig:arch}. It consists of two key components: an Actor that produces SQL statements, and a Critic that verifies if the produced SQL is correct or not. The Critic involves two components: an LLM Critic that verifies the semantic correction of the generated SQL by referring to an LLM, and an Execution Critic that verifies the syntactic correctness of the generated SQL by accessing the database. The LLM Critic is prompted to return True or False, by using the whole information used by the Actor and the generated SQL. The Execution Critic executes the generated SQL via the database API, and returns True if no error is received, otherwise False. Note that the Execution Critic will return True if the SQL is valid in syntax, even though the returned result may be empty due to semantic errors. In contrast, the LLM Critic can find semantic errors as well as syntactic errors. In implementation, we first use the Execution Critic to make sure the SQL is syntactically valid and then ask the LLM Critic to check semantic errors. 

Usually we use the same LLM to play the role of Actor and LLM Critic, though different LLMs are acceptable, e.g., an LLM Critic fine-tuned to the verification task. 
The entire Critic module outputs True if the SQL passes both the LLM Critic and the Execution Critic, otherwise outputs False. If the Critic output is False, the Actor is signified to regenerate the SQL statement. 
The process is repeated until the SQL generated by the Actor is verified to be True by the Critic, or the number of regeneration reaches a pre-defined maximum value. We take the SQL generated by the Actor in the final iteration as the output of the Text-to-SQL system, which means that if the maximum number of regenerations is reached, the regenerated SQL will be output directly without Critic check. Some examples of the solving-and-verifying process are presented in \hyperref[app:sample]{Appendix A}.

\subsection{Performance bound}

A key question is, if the Actor-Critic process assuredly leads to performance improvement. The complete theory requires deeper research, but performance bound under ideal conditions can be discussed. 

First, notice that there is no guarantee that the Critic can offer performance gains. For instance, if the Critic performs very poorly and it always outputs True or False, then it is totally useless. In another, more extreme case, if it always outputs incorrect judgements, then the performance will be fully corrupted. This means that the accuracy of the Critic directly impacts the entire performance. On the other hand, the accuracy of the Actor is also an important factor to consider: a poor Actor needs more careful check than a powerful Actor. Overall, an Actor-Critic system involves complex interaction between an imperfect Actor and an imperfect Critic, and the stochastic behaviors of these two modules determine the performance of the entire system.

Formally, suppose that the Actor is a random generator, that produces SQL statements following a simple binary distribution, e.g., Bernoulli distribution, with a probability $p$ to generate correct SQLs, and $1-p$ to generate incorrect ones. The Critic is an imperfect examiner, and its False Negative Rate (FNR) is $q$, which means that if an incorrect SQL is observed, the Critic returns True with a probability $q$, and False with a probability $1-q$. In addition, it has a False Positive Rate (FPR) $s$, meaning that if the generated SQL is correct, the Critic returns False with a probability $s$, and True with a probability $1-s$. Considering that the maximum number of AC iterations is $z$, the accuracy of the resultant SQL can be rigorously computed, represented as the following theorem, with the derivation presented in \hyperref[app:theorem]{Appendix B}.

\begin{theorem}
\label{th:kappa}
Suppose that the probability that the Actor produces correct SQL statements is $p$, and the FNR and FPR of the Critic are $q$ and $s$ respectively. If the maximum number of AC iterations is $z$, then the expected performance of the AC process, i.e., the probability that the AC-SQL produces correct SQL statements, denoted by $prob$, is as follows:

\begin{equation}
\label{eq:kappa}
prob = p(1-s)\frac{1-[ ps + (1 - p)(1 - q) ]^{z-1}}{1-[ps + (1 - p)(1 - q)]} + p[ ps + (1 - p)(1 - q) ]^{z-1}. 
\end{equation}
\end{theorem}

We have conducted large-scale simulation experiments to validate Eq.~\ref{eq:kappa}. In these simulation experiments, we randomly sampled $p$,
$q$, $s$ and $z$, and simulated the Actor's generation and the Critic's verification behavior with these parameters. We found that the simulation results exactly match the computation by Eq.~\ref{eq:kappa}. Details of the simulation can be found in \hyperref[app:simulation]{Appendix D}.

\subsection{Analysis}

\begin{figure}
    \centering
    \includegraphics[width=\linewidth]{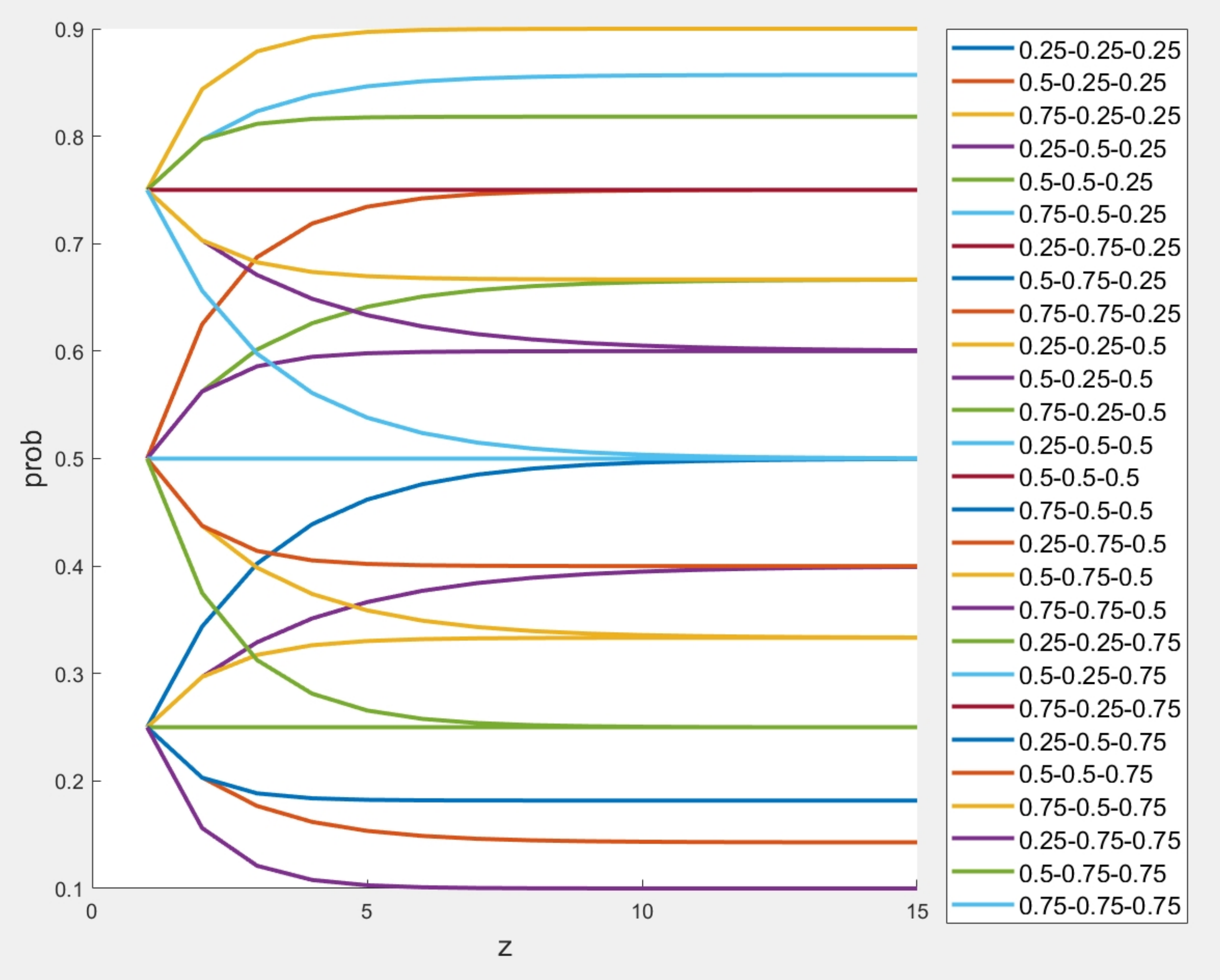}    \caption{The expected performance $prob$ of the AC approach with different settings of parameters. The legend on the right-hand side of the figure shows the value of ($p$, $q$, $s$).}\Description{The influence of the value of ($p$, $q$, $s$) on $prob$}
    \label{fig:z-p}
\end{figure}

\begin{table*}
\centering
\caption{The expected AC performance $prob$ with extreme values of $p$, $q$, $s$, under the condition $ps + (1-p)(1-q) < 1 $.}
\label{tab:special}
\begin{tabular}{l|c|l|c}
\hline
Parameter & Value  &  Meaning & $prob$ when $z\rightarrow \infty$ \\
\hline
\multirow{2}{*}{$p$} & 0 & All the SQLs generated by the Actor are wrong & 0 \\
                     & 1 & All the SQLs generated by the Actor are correct  & 1 \\
\hline
\multirow{2}{*}{$q$} & 0 & All the wrong SQLs are rejected by the Critic & 1 \\
                     & 1 & All the wrong SQLs are accepted by the Critic  & $\frac{p(1-s)}{1-ps}<p$ \\
\hline                     
\multirow{2}{*}{$s$} & 0 & All the correct SQLs are accepted by the Critic & $\frac{p}{p+q(1-p)}>p$ \\
                     & 1 & All the correct SQLs are rejected by the Critic & 0 \\
\hline
\end{tabular}
\end{table*}

\subsubsection{$prob$ with unlimited $z$}
\label{sec:theory:bound}

We first check the expected performance of the AC approach with $z \rightarrow \infty$. Notice that:
\[
ps + (1-p)(1-q) \le p + (1-p)(1-q) \le p + (1 - p) \le 1,
\]
where the equality holds when $p(1-s) = q(p-1)=0$. This condition is equal to $(q=0 \vee p =1) \wedge (p=0) \vee (s=1)$, indicating that at least two of the three parameters $p$, $q$, $s$ are required to be 0 or 1. 

We focus on the practically meaningful situation with $ps+(1-p)(1-q) < 1$, and examine $prob$ with $z \rightarrow  \infty$. A simple computation shows that $prob$, the expected performance of the AC approach is as follows:

\begin{equation}
\label{eq:prob:nn}
prob = \frac{p(1-s)}{p+q-pq -ps}
\end{equation}

\noindent Setting any one of $p,q,s$ to an extreme value of 0 or 1 (the other two between 0 and 1), one can roughly investigate the impact of the three parameters. The results are shown in Table~\ref{tab:special}. 
Firstly notice that if the Actor is perfect or totally fails, the AC approach does not help, which is expected and understandable. For the Critic FNR $q$, a perfect AC performance can be obtained ($prob=1$) when $q=0$, i.e., if it never accepts a wrong SQL. In contrast, if the wrong SQL are all accepted by the Critic, then the AC performance is lower than the base Actor. For the Critic FPR $s$, if $s=0$, the Critic accepts all correct SQL, leading to improved performance. If $q=1$, the Critic works poorly and rejects all correct SQL, making the AC system always outputs wrong SQL. 

It should be highlighted that \emph{for both $q$ and $s$, there is a range of values leading to performance improvement (i.e., better than the base Actor), also a range of values leading to performance drop.} It is easy to find the boundary of the `good' and `bad' ranges is $ s + q = 1$ by solving $\frac{p(1-s)}{p+q-pq -ps} = p$. We summarize the boundary condition as the following corollary.

\begin{corollary}
\label{cor:zinfty}
Let $p$, $q$, $s$, $prob$ defined as in Theorem~\ref{th:kappa}. In the case $z \rightarrow \infty$, $prob \ge p$ if and only if $s+q \le 1$,  and $prob = p$ if $s + q = 1 $ and $p \in (0,1)$. 
\end{corollary}

The derivation from Eq.~\ref{eq:prob:nn} is straightforward. An intuitive interpretation is that if the Critic performs better than randomness overall, then the AC approach will improve the performance.

\subsubsection{$prob$ with limited $z$}

We further investigate the expected performance $prob$ of an AC system with a limited $z$, as shown in Fig.~\ref{fig:z-p}, where we randomly sample some values for $p$,$q$,$s$, and observe the value of $prob$. It can be observed that as $z$ increases, $prob$ tends to stabilize, and when $z > 10$, the value almost converges. Moreover, different combinations of $(p,q,s)$ lead to different consequences, and involving the Critic does not necessarily lead to performance gains. This confirms the analysis in the previous section that it is the interaction among the three parameters that determines the performance, while the new finding is that the impact of the AC approach with a limited $n$ and unlimited $n$ is the same, i.e., if $(p,q,s)$ offers performance gains, it will always do so with any $n$, and a larger $n$ leads to more gains. 

This result can be derived from Theorem~\ref{th:kappa} rigorously, and formally represented as the following corollary. Refer to \hyperref[app:boundary]{Appendix C} for the proof.

\begin{corollary}
\label{cor:z}
Let $p$, $q$, $s$, $prob$ defined as in Theorem~\ref{th:kappa}, $prob \ge p$ if and only if $s+q \le 1$ for any $z$,  and $prob = p$ holds when $s + q = 1$, conditioned on $z >1$ and $p \in (0,1)$. 
\end{corollary}

\subsubsection{$prob$ on $q$-$s$ plane}

\begin{figure}
    \centering
    \includegraphics[width=\linewidth]{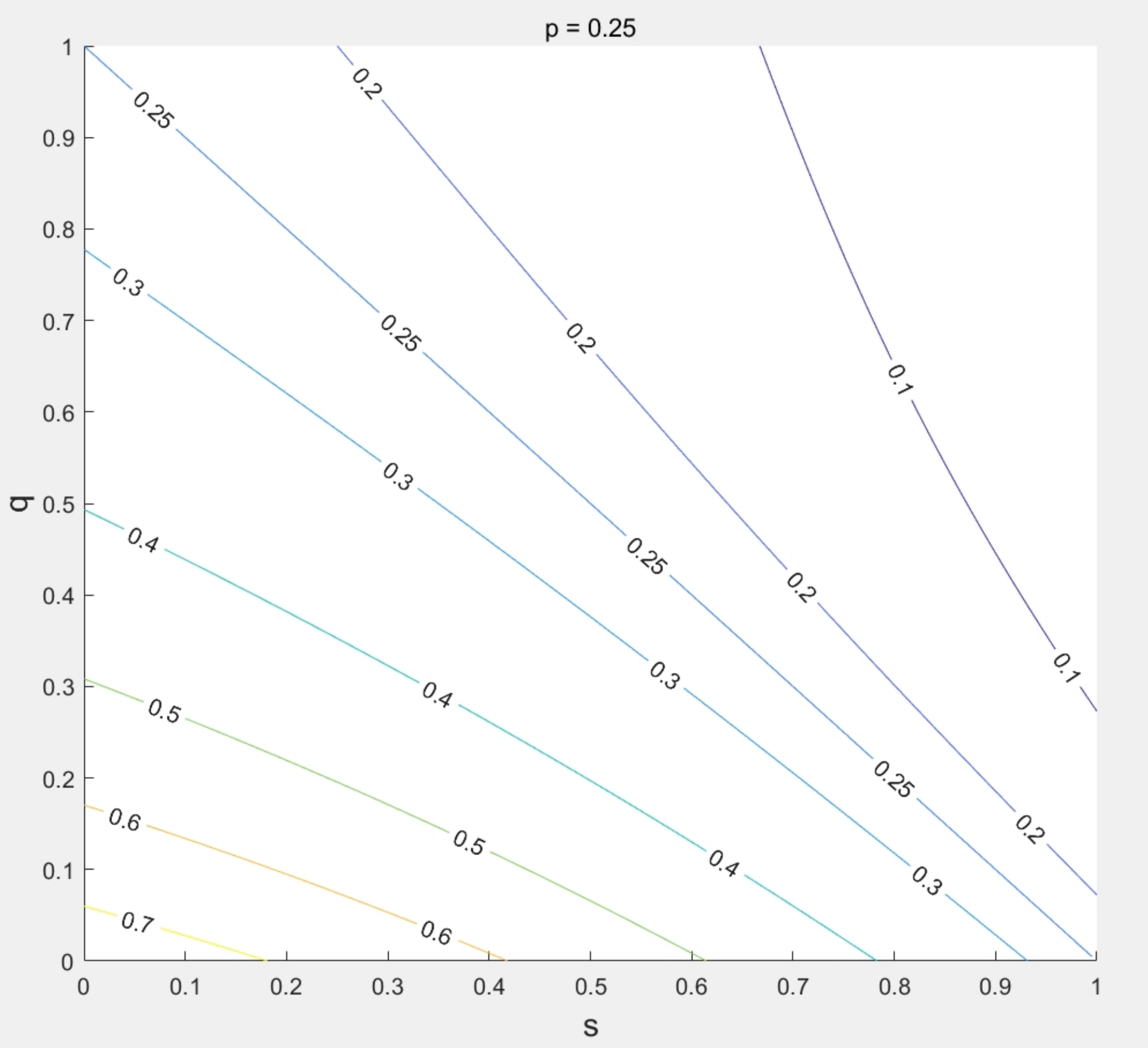}    \caption{$prob$ values on the $q$-$s$ plane, with $p = 0.25$ and $z = 5$.}
    \label{fig:p=0.25}
    \Description{Prob on q-s}
\end{figure}

\begin{figure}
    \centering
    \includegraphics[width=\linewidth]{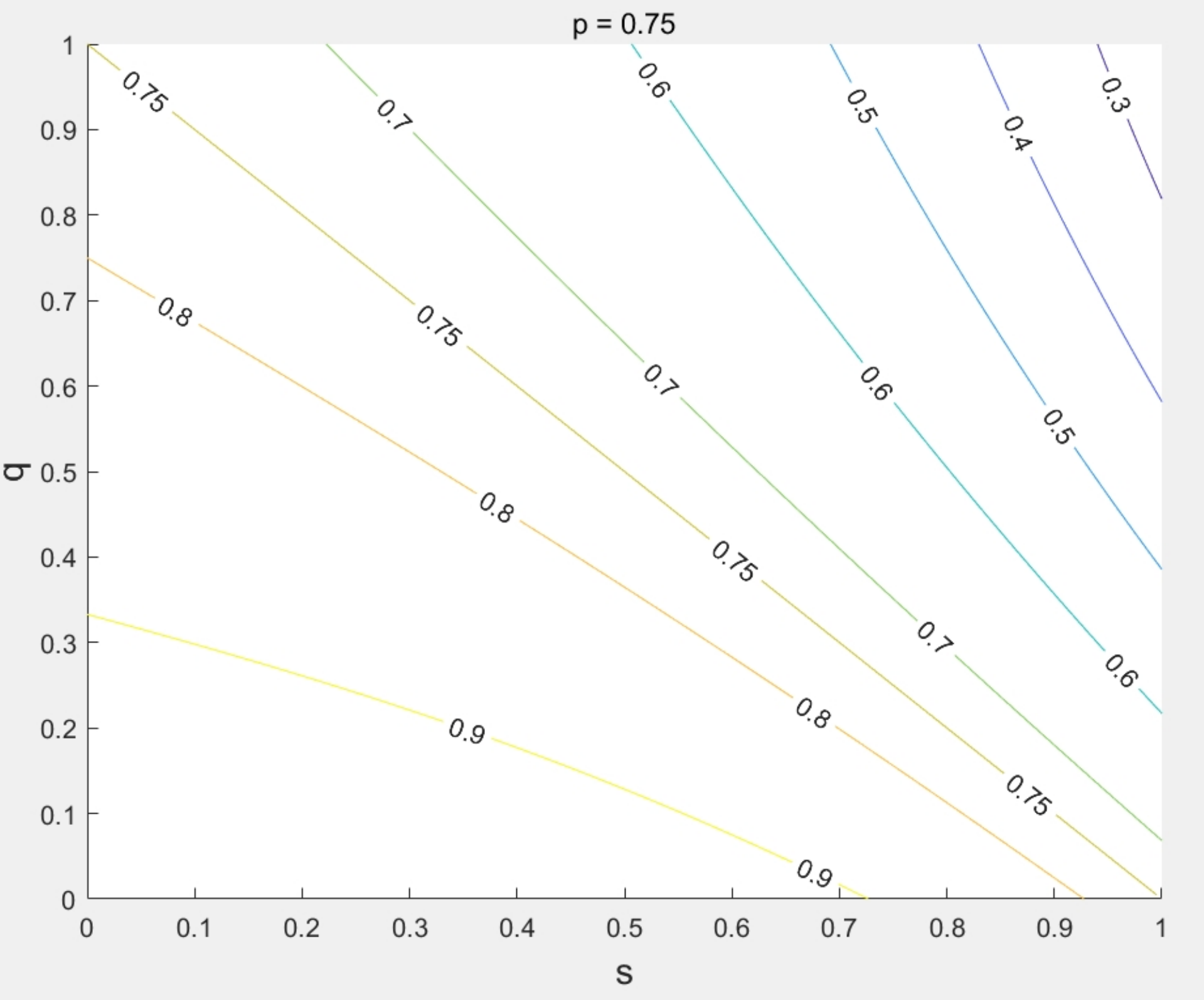}    \caption{$prob$ values on the $q$-$s$ plane, with $p = 0.75$ and $z = 5$.}
    \label{fig:p=0.75}
    \Description{Prob on q-s}
\end{figure}

In most cases, we are more concerned about the question `Which Critic can improve performance with a fixed Actor'. 
In the previous analysis, we have established the boundary condition $q+s=1$, however a global picture of the performance gains/losses that the AC approach can yield is also useful. To gain that global picture, we plot the $prob$ contour in the $q-s$ plane with $z = 5$, as presented in Fig.~\ref{fig:p=0.25} for the case $p=0.25$ and Fig.~\ref{fig:p=0.75} for the case $p=0.75$. It can be seen that smaller $q$ and $s$ lead to better performance, and at the origin area, the performance could be rather good although the base performance $p$ is low. In contrast, if $q$ and $s$ are large, the Critic performs poorly and it will degrade performance rather than offer improvement. 

These figures also show the boundary of `good range' and `bad range' of $(q,s)$. According to Corollary~\ref{cor:z}, this boundary is represented by the condition $s + p = 1$, which is the anti-diagonal line from the top-left corner (0,1) and the bottom-right corner (1,0). All the settings of ($q,s$) on this line lead to $prob=p$.
Specifically, the point (0,1) means =`accept all', and (1,0) means `reject all'. Clearly $prob=p$ holds in these two special cases. According to Corollary~\ref{cor:z}, the settings of ($q,s$) below this boundary result in performance improvement, while the settings above this boundary degrade the performance.

Overall, the above observations indicate that even although the Actor is weak and cannot generate solutions with a high accuracy, we still can expect a high performance with the AC approach if the Critic is strong. \textbf{In particular, if the Critic performs not worse than random decision, the AC approach will surely outperform the Actor. }
This means that the AC approach is safe and reliable, an attractive property compared to counterpart methods.

\section{Experiments}

\subsection{Datasets} 

We conducted experiments on the Spider-dev dataset~\cite{yu2018spider} and two related datasets Spider-DK~\cite{gan2021exploring} and Spider-SYN~\cite{gan2021towards}. Note that we focus on the zero-shot paradigm and so all the datasets are used for testing, i.e., no training or demonstration samples are required. More details of the three datasets are presented below. 

\begin{itemize}
\item \textbf{Spider-dev\footnote{\url{https://yale-lily.github.io/spider}}}: Spider is a large-scale, complex and cross-domain text-to-SQL dataset, consisting of 10,181 questions and 5,693 SQL questions on 200 databases, covering 138 different domains. Spider-dev is the develop set of Spider, containing 1,034 queries. 

\item \textbf{Spider-DK\footnote{\url{https://github.com/ygan/SpiderDK}}}: Spider-DK is a variant of Spider-dev designed to test queries that demand domain knowledge, making them more challenging. It was created by selecting and modifying samples from Spider-dev to ensure that text-to-SQL inference requires a deeper understanding of domain-specific knowledge. The dataset comprises 535 text-SQL pairs, with 270 pairs identical to the original Spider samples, while the remaining 265 pairs have been adjusted to make sure domain knowledge is required for accurate inference. Note that the size of Spider-DK is smaller than Spider-dev, as it only involves the queries that require domain knowledge. 

\item \textbf{Spider-SYN\footnote{\url{https://github.com/ygan/Spider-Syn}}}: 
Spider-SYN is a variant of Spider-dev, designed to test queries with expression variations. To achieve this, the queries in Spider-dev were modified by replacing schema-related words with manually selected synonyms that reflect real-world question paraphrases. Spider-SYN and Spider-dev are the same size, i.e., 1,034 queries.
\end{itemize}

\subsection{Models and settings} 

\begin{table*}
\centering
\caption{Properties of LLMs tested in the experiment.}
\label{tab:model}
\begin{tabular}{l|p{0.8\linewidth}}
\hline
Model & Description\\
\hline
\textbf{LLaMA2}~\cite{touvron2023llama2} & A collection of retrained and fine-tuned open-source language models, introduced by Meta in July 2023. It is one of the most advanced and impactful in terms of performance and community contribution. We conduct our evaluations on LLaMA2-7B and LLaMA2-13B. \\
\hline
\textbf{LLaMA3}~\cite{llama32024} & A collection of pretrained and fine-tuned open-source language models, introduced by Meta in April 2024. It represents the latest advancements in the LLaMA series of large language models. We use LLaMA3-8B in our experiments.  \\
\hline
\textbf{Guanaco}~\cite{dettmers2024qlora} & A series of open-source models fine-tuned from the LLaMA models with the OASST1 dataset, through 4-bit QLoRA tuning. It was introduced by University of Washington in April 2023. It claims to achieve 99.3\% of ChatGPT performance with only 24 hours of fine-tuning on a single GPU. We experience with Guanaco-7B and Guanaco-33B.\\
\hline
\textbf{Vicuna}~\cite{vicuna2023}  & A series of open-source models fine-tuned from LLaMA models with supervised instruction fine-tuning, introduced by a collaborative effort involving multiple institutions in March 2023. It claims to achieve 90\% ChatGPT quality based on an automated evaluation with GPT-4. We test the 7B, 13B and 33B Vicuna models. \\
\hline
\textbf{Gemma}~\cite{team2024gemma} & A family of open-source models introduced by Google in February 2024. Compared to other models, Gemma is a powerful but lightweight model that can be deployed on a regular laptop. We select Gemma-7B in our experiments. \\
\hline
\textbf{GPT-3.5}~\cite{gpt35} & An advanced commercial language model introduced by OpenAI in November 2022. It is the most popular model and is frequently used as a baseline in various experiments. We use the version gpt-3.5-turbo-0125.\\
\hline
\textbf{GPT-4o}~\cite{gpt4o}  & The most powerful commercial model developed by OpenAI so far. We use the version gpt-4o-2024-05-13.\\
\hline
\end{tabular}
\end{table*}

We experimented with 4 series of open source models (LLaMA, Guanaco, Vicuna, Gemma) and 2 commercial models (GPT-3.5 and GPT-4o). The properties of these models are summarized in Table~\ref{tab:model}. For the open-source models, we downloaded them and ran them on a server with four Nvidia RTX 3090 GPUs. For the commercial models, we used the online APIs to access the OpenAI service. 
All the open-source models were downloaded from \href{https://huggingface.co/}{Hugging Face}. 

For the AC-SQL method, the maximum number of regenerations is set to be 5, considering the trade-off between run-time speed and accuracy. For the execution Critic, we used the SQLite interface to access the databases. In all the AC-SQL experiments, we used the same LLM as the Actor and the Critic, although there is no such limit in both theory and practice.

\subsection{Prompts}

We chose simple prompts to instruct an LLM to play the role of Actor and LLM Critic. The prompt for the Actor is exemplified in List~\ref{list:ocr}. It simply informs the LLM SQL statements used to create the schema, and asks the model to create a SQL statement for the query without explanation. 

\begin{lstlisting}[caption={Example of the Actor Prompt}, label=list:ocr, xleftmargin=\parindent, xrightmargin=\parindent]
<@\textcolor{red}{CREATE TABLE continents(ContId int primary key ,\\
Continent text ,foreign key(ContId) references\\
countries(Continent));CREATE TABLE countries\\
(CountryId int primary key ,CountryName text ,\\Continent
int ,foreign key(Continent) references \\
continents(ContId));}@>

<@\textcolor{blue}{Create a SQL query only for the given questions using database schema above without explanation:How many continents are there?}@>
\end{lstlisting}

The prompt for the LLM Critic is demonstrated in List~\ref{list:critic}. It provides the LLM all the information used by the Actor, i.e., the SQL statements used in schema creation and the question in natural language, plus the SQL statement produced by the Actor. It then asks the LLM to issue True if it believes the Actor-generated SQL being correct and False otherwise.

\begin{lstlisting}[caption={Example of the Critic Prompt}, label=list:critic, xleftmargin=\parindent, xrightmargin=\parindent]
<@\textcolor{red}{CREATE TABLE continents(ContId int primary key ,\\
Continent text ,foreign key(ContId) references\\
countries(Continent));CREATE TABLE countries\\
(CountryId int primary key ,CountryName text ,\\
Continent int ,foreign key(Continent) references continents(ContId));}@>

<@\textcolor{blue}{Answer True if the SQL query is correct and False if incorrect without explanation. Question: What are the death and injury situations caused by the ship with tonnage 't'? SQL: SELECT killed, injured FROM death WHERE caused\_by\_ship}@>
\end{lstlisting}

\subsection{Main results} 

The execution accuracy (EX) was used as the evaluation metric. It assesses the correctness of the produced SQL by comparing the execution results of the produced SQL with the ground-truth SQL~\cite{zhong2020semantic}. Note that it is likely that two SQL statements are different literally but produce the same results. This means EX focuses on results returned by executing the SQL statement, rather than its text form. 

Table~\ref{tab:result} reports the results on Spider-dev with and without the Actor-Critic process. We report the EX results and absolute EX improvement. We also report relative reduction on error rates, where we define the error rate as (1 - EX). From the results, it can be observed that with all the models, the AC-SQL approach offers remarkable EX improvement and the relative error rate reduction is rather significant. Even with the strong GPT-4o model, the AC-SQL approach yields an improvement of 5.6 percentage points in EX and an error rate reduction of 25.1\%. This consistent and remarkable performance improvement demonstrates that the AC-SQL approach is highly effective despite its incredible simplicity.

\begin{table}
\centering
\caption{Main Results on Spider-dev with and without Actor-Critic, denoted by AC(+) and AC(-) respectively. The column Abs(\%) reports the absolute improvement in execution accuracy; Rel.(\%) reports the relative error reduction. }
\label{tab:result}
\resizebox{\linewidth}{!}{
\begin{tabular}{l|c|c|c|c|c|c}
\hline
Model &  Open & Context &  \multicolumn{2}{c|}{EX (\%)} & Abs.(\%) & Rel.(\%)\\
\hline
\multicolumn{3}{c|}{} & AC (-) & AC (+) & \multicolumn{2}{c}{}\\
\hline
\href{https://huggingface.co/meta-llama/Llama-2-7b-hf}{LLaMA2-7B} 
& $\checkmark$& 4K & 1.2 & 10.9 &  9.7 & 10.9\\
\href{https://huggingface.co/meta-llama/Llama-2-13b-hf}{LLaMA2-13B} 
& $\checkmark$& 4K & 5.0 & 25.0 &  20.0 & 21.1\\
\href{https://huggingface.co/meta-llama/Meta-Llama-3-8B-Instruct}
{LLaMA3-8B} 
& $\checkmark$& 8K & 32.6 & 67.7 & \textbf{ 35.1} & 52.1\\
\href{https://huggingface.co/timdettmers/guanaco-7b}{Guanaco-7B} 
& $\checkmark$& 4K & 0.9 & 11.3 &  10.4 & 10.5\\
\href{https://huggingface.co/timdettmers/guanaco-33b-merged}{Guanaco-33B} 
& $\checkmark$& 4K & 28.3 & 37.3 &  9.0 & 12.6\\
\href{https://huggingface.co/lmsys/vicuna-7b-v1.5-16k}{Vicuna-7B} 
& $\checkmark$& 16K & 15.3  & 20.6 &  5.3 & 6.3\\
\href{https://huggingface.co/lmsys/vicuna-13b-v1.5-16k}{Vicuna-13B} 
& $\checkmark$& 16K & 21.7& 32.4 &  10.7 & 13.7\\
\href{https://huggingface.co/lmsys/vicuna-33b-v1.3}{Vicuna-33B} 
& $\checkmark$& 2K & 43.7 & 61.0 &  17.3 & 30.7\\
\href{https://huggingface.co/google/gemma-7b-it}{Gemma-7B} 
& $\checkmark$& 8K & 30.3 & 65.3 &  35.0 & 50.2\\
\hline
{GPT-3.5} 
& $\times$& 16K & 58.8 & 70.4 &  11.6 & 28.2\\
{GPT-4o} 
& $\times$& 128K & 72.1 & 77.7 &  5.6 & 25.1\\
\hline
\end{tabular}
}
\end{table}

Additionally, we extended the experiments to the two more challenging datasets:  Spider-DK and Spider-SYN.  The results are shown in Table~\ref{tab:result-db}. It can be observed that the AC-SQL approach generalizes well across all the datasets, and in all the tested cases, the improvement is highly significant, demonstrating the generality of the proposed method.

Perhaps the most interesting observation is that in all the test cases, there is no performance reduction, even with the rather poor model LLaMA2-7B. At the first glance this is somewhat surprising as a very weak Critic may cause performance loss rather than gains. However, according to the boundary condition $q+s=1$ shown in Section~\ref{sec:theory:bound}, if the Critic performs better than random acceptance/rejection, then performance gain will be guaranteed. Although LLaMA2-7B is weak in SQL generation, it is very likely that it can perform better than randomness in the verification task, at least when combined with database API. This is understandable as the weak LLM still learned rich knowledge and the verification  task is simpler than the generation task. 

We also tested if the expected performance derived from Theorem~\ref{th:kappa} matches the true performance of a real system. To perform the test, we calculated the values of $p$, $q$, and $s$ for the Vicuna-33B model on the Spider dataset, resulting \(p=0.3774\), \(q=0.2541\), and \(s=0.1973\) . Setting $z = 5$, Theorem~\ref{th:kappa} predicts that the AC-SQL performance is $prob = 0.6334$. This result closely matches our experimental outcome, which showed an accuracy of 61.0\%. This indicates that Theorem~\ref{th:kappa} can roughly predict the behavior of the AC-SQL approach.

 \begin{table}
\centering
\caption{Results on Spider-DK and Spider-SYN with and without Actor-Critic, denoted by AC(+) and AC(-) respectively. }
\label{tab:result-db}
\begin{tabular}{l|c|c|c|c}
\hline
 &  \multicolumn{4}{|c}{EX (\%)} \\ 
\hline
&  \multicolumn{2}{|c}{Spider-DK} & \multicolumn{2}{|c}{Spider-SYN}\\
\hline
Model &   AC (-) & AC (+) & AC (-) & AC (+) \\
\hline
{LLaMA2-7B} 
& 1.2 & 5.2 &  1.6 & 6.5\\
{LLaMA2-13B} 
& 10.2 & 23.7 &  6.2 & 18.8\\
{LLaMA3-8B} 
& 30.1 & 57.8 & 28.9 & 55.9\\
{Guanaco-7B} 
& 2.7 & 11.1 & 1.5 &  8.6\\
{Guanaco-33B} 
& 26.7 & 33.7 & 24.4 &  30.1\\
{Vicuna-7B} 
& 11.8 & 15.6 & 9.2 &  13.3\\
{Vicuna-13B} 
& 18.8 & 25.1 & 15.3 &  20.7\\
{Vicuna-33B} 
&33.2& 37.2 & 32.6 & 40.0 \\
{Gemma-7B} 
& 23.2 & 58.2 & 20.1 &  51.2\\
\hline
{GPT-3.5} 
& 62.5 & 64.6 & 64.6 &  66.0\\
{GPT-4o} 
& 67.0 & 68.2 & 65.8 &  68.5\\
\hline
\end{tabular}
\end{table}

\subsection{Ablation study}

In this section, we experiment with the individual contributions of the LLM Critic and the Execution Critic. The results are shown in Table~\ref{tab:ablation:critic}. It can be seen that, except for the Vicuna-7B and Vicuna-13B models, both the LLM Critic module and the Execution Critic module significantly improve the model's performance, and combining the two Critics leads to further performance gain. However, there is not a consistent observation which Critic contributes more, and how significant when they are combined. For instance, with LLaMA3-8B, the LLM Critic is much more effective than the Execution Critic (60.6 vs 48.0), and their combination is also effective (67.7). This means that the LLM Critic detected a large body of semantic errors thus complementing with the Execution Critic that focuses on detecting syntax errors. In contrast, for the most powerful GPT-4o, the two Critics lead to similar performance (77.3 vs 77.1), and the additional gains obtained from their combination offers is marginal (77.7). This indicates that the LLM-Critic detected more syntax errors than semantic errors, and thus played a similar role as the Execution Critic. 

For Vicuna-7B and Vicuna-13B, the Execution Critic is highly effective while the LLM Critic leads to performance degradation. Analysis shows that the $(q,s)$ pair of the Critic with these two models is (0.68,0.55) and (0.60, 0.41), respectively. This means that the Critic really worked worse than random decision in the two tests. We attribute this seemingly `unlikely' situation to the simple, even something shabby prompt used to invoke the Critic in our study.

\begin{table}
\centering
\caption{Results on Spider-dev without any Critic (AC(-)), with LLM Critic (AC-LLM), with Execution Critic (AC-Ex), or with both Critics, i.e., the standard AC-SQL (AC(+)). }
\label{tab:ablation:critic}
\begin{tabular}{l|c|c|c|c}
\hline
&  \multicolumn{4}{|c}{EX (\%)} \\
\hline
Model      & AC (-) & AC-LLM  & AC-Ex & AC (+)  \\
\hline
{LLaMA2-7B} & 1.2 & 8.4 & 8.3 & 10.9 \\
{LLaMA2-13B} & 5.0 & 23.1 & 23.7 & 25.0 \\
{LLaMA3-8B} & 32.6 & 60.6 & 48.0 & 67.7 \\
{Guanaco-7B} & 0.9 & 6.5 & 6.4 & 11.3 \\
{Guanaco-33B} & 28.3 & 33.3 & 36.2 & 37.3 \\
{Vicuna-7B} & 15.3 & 7.6 & 35.9 & 20.6 \\
{Vicuna-13B} & 21.7 & 18.1 & 47.1 & 32.4 \\
{Vicuna-33B} & 43.7 & 47.7 & 48.7 & 61.0 \\
{Gemma-7B} & 29.9 & 59.4 & 63.7 & 65.3 \\
\hline
{GPT-3.5} & 58.8 & 66.5 & 65.1 & 70.4 \\
{GPT-4o} & 72.1 & 77.3 & 77.1 & 77.7 \\
\hline
\end{tabular}
\end{table}

\section{Limitations}

The Critic in this study responds in a simple form of "True" or "False". This simplicity is crucial for us to establish the AC theory. However, more sophisticated responses involve richer information and thus make use of LLMs' potential more effectively. Some previous studies have investigated using complex responses, e.g,~\cite{pourreza2024din,dong2023c3}. A key problem of these methods, however, is that the theoretical guarantee for performance gains will be lost. Establishing theoretical foundation for complex critics is an inspiring direction. 

Another limitation of the AC approach lies in the execution efficiency. Since it requires an iterative process, the run-time efficiency is low. How to improve inference speed is a central problem. 

Finally, we used the same LLM as the Actor and Critic in this study, which is essentially `self checking'. Although generation and verification are two different capacities and not fully dependent with each other, self-checking is probably not the best way, just like human beings tend to overlook their own shortcomings. We need to investigate a `peer review' strategy that uses different LLMs to take the role of Actor and Critic independently.

\section{Conclusion}

This paper proposes a simple yet effective Text-to-SQL method called AC-SQL. It designs two agents, an Actor and a Critic, playing the roles of SQL generation and verification respectively. Theoretical analysis shows that the performance is ensured to be improved only if the Critic behaves better than random acceptance/rejection. Comprehensive experiments conducted on a broad range of models and datasets demonstrated remarkable performance improvement with this simple approach. 

According to the AC theory, the performance of the Critic is crucial. Interestingly, constructing a stronger Critic is supposed to be much easier than a stronger Actor. We need to explore how to train a more `criticizing' LMM as the Critic. Mechanisms like democratic voting are also interesting: rather than a single Critic, a critic panel may lead to further improvement.

%\section{Acknowledgments}
%\begin{acks}
%  We would like to express our deepest gratitude to our colleagues for their invaluable assistance during the experimental phase of this research. We are also grateful to our lab at the university for providing the necessary resources and facilities. Their support and contributions were instrumental in the successful completion of this study.
%\end{acks}

\bibliographystyle{ACM-Reference-Format}
\bibliography{reference}
\small

%%%%%%%%%%%%%%%%%%%%%%%%%%%%%%%%%%%%%%%%%%%%%%%%%%%%%%%%%%%%
\appendix

\section{Examples of AC-SQL Iteration}
\label{app:sample}

\noindent \textbf{Example 1:}

\noindent \colorbox{yellow}{User to Actor and Critic}: 
"CREATE TABLE battle ( id INT, name TEXT, date TEXT, bulgarian\_commander TEXT, latin\_commander TEXT, result TEXT, PRIMARY KEY ( id ) ); 

CREATE TABLE death ( caused\_by\_ship\_id INT, id INT, note TEXT, killed INT, injured INT, PRIMARY KEY ( id ), FOREIGN KEY ( caused\_by\_ship\_id ) REFERENCES ship (id) ); 

CREATE TABLE ship ( lost\_in\_battle INT, id INT, name TEXT, tonnage TEXT, ship\_type TEXT, location TEXT, disposition\_of\_ship TEXT, PRIMARY KEY ( id ), FOREIGN KEY ( lost\_in\_battle ) REFERENCES battle (id) );" 

\colorbox{yellow}{User to Actor}: 
"Create a SQL query only for the given questions using database schema above without explanation: What are the death and injury situations caused by the ship with tonnage 't '?"

\colorbox{yellow}{Actor}: "SELECT killed, injured FROM death WHERE caused\_by\_ship\_id = t;"

\colorbox{yellow}{User to Critic}: "Answer True if the SQL query is correct and False if incorrect without explanation,question: What are the death and injury situations caused by the ship with tonnage 't ' ?  SQL: SELECT killed, injured FROM death WHERE caused\_by\_ship\_id = t”

\colorbox{yellow}{Critic}: "False"

\colorbox{yellow}{User to Actor}: "Please provide a new SQL query to the question only without explanation: What are the death and injury situations caused by the ship with tonnage 't ' ?"

\colorbox{yellow}{Actor}: "SELECT T1.killed , T1.injured FROM death AS T1 JOIN ship AS t2 ON T1.caused\_by\_ship\_id = T2.id WHERE T2.tonnage = 't'"

\colorbox{yellow}{User to Critic}: "Answer True if the SQL query is correct and False if incorrect without explanation,question: What are the death and injury situations caused by the ship with tonnage 't ' ?  SQL: SELECT T1.killed , T1.injured FROM death AS T1 JOIN ship AS t2 ON T1.caused\_by\_ship\_id = T2.id WHERE T2.tonnage = 't'"

\colorbox{yellow}{Critic}: "True"

\colorbox{yellow}{Output SQL}: " SELECT T1.killed , T1.injured FROM death AS T1 JOIN ship AS t2 ON T1.caused\_by\_ship\_id = T2.id WHERE T2.tonnage = 't'"

\noindent  \textbf{Example 2:}

\noindent  \colorbox{green}{User to Actor and Critic}: 
"CREATE TABLE battle ( id INT, name TEXT, date TEXT, bulgarian\_commander TEXT, latin\_commander TEXT, result TEXT, PRIMARY KEY ( id ) ); 

CREATE TABLE death ( caused\_by\_ship\_id INT, id INT, note TEXT, killed INT, injured INT, PRIMARY KEY ( id ), FOREIGN KEY ( caused\_by\_ship\_id ) REFERENCES ship (id) ); 

CREATE TABLE ship ( lost\_in\_battle INT, id INT, name TEXT, tonnage TEXT, ship\_type TEXT, location TEXT, disposition\_of\_ship TEXT, PRIMARY KEY ( id ), FOREIGN KEY ( lost\_in\_battle ) REFERENCES battle (id) );" 

\colorbox{green}{User to Actor}: "Create a SQL query only for the given questions using database schema above without explanation: What are the death and injury situations caused by the ship with tonnage 't '?"

\colorbox{green}{Actor}: "SELECT killed, injured FROM death WHERE caused\_by\_ship\_id = t;"

\colorbox{green}{User to Critic}: "Answer True if the SQL query is correct and False if incorrect without explanation,question: What are the death and injury situations caused by the ship with tonnage 't ' ?  SQL: SELECT killed, injured FROM death WHERE caused\_by\_ship\_id = t"

\colorbox{green}{Critic}: "False"

\colorbox{green}{User to Actor}: "Please provide a new SQL query to the question only without explanation: What are the death and injury situations caused by the ship with tonnage 't ' ?"

\colorbox{green}{Actor}: "SELECT killed, injured FROM death WHERE caused\_by\_ship\_id = t;"

\colorbox{green}{User to Critic}: "Answer True if the SQL query is correct and False if incorrect without explanation,question: What are the death and injury situations caused by the ship with tonnage 't ' ? 

SQL: SELECT killed, injured FROM death WHERE caused\_by\_ship\_id = t”

\colorbox{green}{Critic}: "False"

\colorbox{green}{User to Actor}: "Please provide a new SQL query to the question only without explanation: What are the death and injury situations caused by the ship with tonnage 't ' ?"

\colorbox{green}{Actor}: "SELECT killed, injured FROM death WHERE caused\_by\_ship\_id = t;"

\colorbox{green}{User to Critic}: "Answer True if the SQL query is correct and False if incorrect without explanation,question: What are the death and injury situations caused by the ship with tonnage 't ' ?  

SQL: SELECT killed, injured FROM death WHERE caused\_by\_ship\_id = t"

\colorbox{green}{Critic}: "False"

\colorbox{green}{Output SQL}:"SELECT killed, injured FROM death WHERE caused\_by\\\_ship\_id = t;"

\section{Derivation of Theorem \ref{th:kappa}}
\label{app:theorem}

% In this section, we present the proof of Theorem \ref{th:kappa}.

% More specifically,when  the maximum number of AC iterations is z, the final correct result can be divided into two situations:

% 1. In the i-th (i<z) iteration, the correct result was obtained. The condition for this is that the critic returned False in the first i-1 iterations, the actor generated the correct result in the i-th loop, and the critic returned True.

% 2. The correct result was obtained in the last iteration. The condition for this is that the actor generated the correct result.

% Therefore, the probabilities of these two situations should be calculated separately and then summed up.

\begin{proof}
According to the definition, the Actor generates correct SQLs with a probability $p$; the Critic returns True with a probability $q$ for an incorrect SQL, and returns False with a probability $s$ for a correct SQL. The maximum number of AC iterations is $z(z > 1)$. 
Let $C_i = T/F$ denote the returned value of the Critic in the $i$-th iteration, and $A_i = T/F$ denote the correctness of the SQL statement generated in the $i$-th iteration.  The AC-SQL outputs a correct SQL in two possible scenarios: 

\textbf{Scenario 1: The Critic returns True at the $i$-th iteration ($i<z$) while returns False before, and the returned SQL is correct.} 

\begin{align}
& P(Scenario1, i) \nonumber\\ 
&= P(C_{1,2,...,i-1} = F, C_i = T, A_i = T, i < z) \nonumber\\ 
&= P(C_{1,2,..., i-1} = F, i < z) \cdot P(C_i = T, i < z) \cdot P(A_i = T, i < z) \nonumber\\
&=  P(C_j = F, j = 1,..., i-1, i < z) ^{i-1} \cdot P(C_i = T) \cdot P(A_i = T) \nonumber \\
&= [ P(A_j = T, C_j = F, j = 1,..., i-1, i < z) \nonumber\\
&+ P(A_j = F, C_j = F, j = 1,..., i-1, i < z) ]^{i-1}\nonumber\\
& \cdot P(C_i = T) \cdot P(A_i = T) \nonumber\\
&= p(1 - s)[ ps + (1 - p)(1 - q) ]^{i-1}\nonumber 
\end{align}

\textbf{Scenario 2: The Critic always returns False until the AC process reaches the maximum number of iterations, and the returned SQL is correct.} 

\begin{align}
& P(\text{Scenario2}) \nonumber \\
&= P(C_{1,2,...,z-1} = F, A_z = T) \nonumber\\ 
&= P(C_{1,2,..., z-1} = F) \cdot P(A_z = T) \nonumber\\
&=  P(C_j = F, j = 1,..., z-1) ^{z-1} \cdot P(A_z = T) \nonumber\\
&= [ P(A_j = T, C_j = F, j = 1,..., z-1) \nonumber\\
&+ P(A_j = F, C_j = F, j = 1,..., z-1) ]^{z-1} \cdot P(A_z = T) \nonumber \\
&= p[ ps + (1 - p)(1 - q) ]^{z-1}\nonumber 
\end{align}

Accumulating the probabilities from Scenario1 and Scenario2 results in the probability that the AC process produces correct SQLs: 

\begin{align}
Prob &= \sum_{i=1}^{z-1}P(\text{Scenario1, i}) + P(\text{Scenario2}) \nonumber\\
&= \sum_{i=1}^{z-1}p(1 - s)[ ps + (1 - p)(1 - q) ]^{i-1}+ p[ ps + (1 - p)(1 - q) ]^{z-1}\nonumber\\
&= p(1-s)\frac{1-[ ps + (1 - p)(1 - q) ]^{z-1}}{1-[ps + (1 - p)(1 - q)]} + p[ ps + (1 - p)(1 - q) ]^{z-1} \nonumber
\end{align}

\end{proof}

\begin{table*}
\centering
\caption{AC-SQL performance derived from Theorem~\ref{th:kappa} and obtained from simulation experiments, in the form of (Theoretical, Simulation).}
\label{tab:the and exp}
\begin{tabular}{c|c|c|c|c|c}
\hline
 &  \multicolumn{5}{|c}{z} \\ 
\hline
   (p,q,s) & 1 & 2 & 3 & 4 & 5  \\
\hline
(0.25,0.25,0.25) & (0.25000,0.25013) & (0.34375,0.34381) & (0.40234,0.40232) & (0.43896,0.43903) & (0.46185,0.46187)\\
(0.25,0.25,0.75) & (0.25000,0.25004) & (0.25000,0.24997) & (0.25000,0.25004) & (0.25000,0.24993) & (0.25000,0.24998)\\
(0.25,0.75,0.25) & (0.25000,0.25002) & (0.25000,0.24999) & (0.25000,0.24998) & (0.25000,0.24995) & (0.25000,0.24994)\\
(0.25,0.75,0.75) & (0.25000,0.24999) & (0.15625,0.15629) & (0.12109,0.12105) & (0.10791,0.10794) & (0.10297,0.10299)\\
(0.75,0.25,0.25) & (0.75000,0.75002) & (0.84375,0.84372) & (0.87891,0.87892) & (0.89209,0.89209) & (0.89703,0.89706)\\
(0.75,0.25,0.75) & (0.75000,0.75008) & (0.75000,0.74993) & (0.75000,0.74999) & (0.75000,0.74998) & (0.75000,0.74998)\\
(0.75,0.75,0.25) & (0.75000,0.74999) & (0.75000,0.75009) & (0.75000,0.74995) & (0.75000,0.75001) & (0.75000,0.74999)\\
(0.75,0.75,0.75) & (0.75000,0.74997) & (0.65625,0.65625) & (0.59766,0.59767) & (0.56104,0.56101) & (0.53815,0.53818)\\
\hline
  \end{tabular}
\end{table*}

\section{Derivation of Corollary \ref{cor:z}}
\label{app:boundary}

\begin{proof}

For $z >1$ and $p \in (0,1)$, we start with the given inequality \(q + s \leq 1\):

\begin{equation}
\begin{aligned}
& q + s \le 1 \\
\leftrightarrow &1-q \ge s \\
\leftrightarrow &(1-p)(1-q) \ge (1-p)s \\
\leftrightarrow &ps + (1-p)(1-q) \ge s \\
\end{aligned} \nonumber
\end{equation}

Let $A = ps + (1-p)(1-q)$. It is easy to verify that  $s \leq A \leq 1 $, implying $1-A \geq 0$ and $A^{z-1} \geq 1$. Obviously, the equality holds if and only if $s = 1$ and $q = 0$, and in this case $prob = p$ as all SQLs are rejected until AC reaches the maximum number of iterations.

For $s \neq 1$ and $q \neq 0$, we have $1-A > 0$ and $A^{z-1} < 1$.
Therefore:

\begin{equation}
\begin{aligned}
& q + s \le 1 \\
\leftrightarrow &A \ge s \\
\leftrightarrow &1-s \ge 1 - A \\
\leftrightarrow &\frac{1-s}{1-A} \ge 1 \\
\leftrightarrow &\frac{1-s}{1-A}(1-A^{z-1}) \ge 1-A^{z-1}\\
\leftrightarrow &(1-s)\frac{1-A^{z-1}}{1-A}+A^{z-1} \ge 1\\
\leftrightarrow &p(1-s)\frac{1-A^{z-1}}{1-A}+pA^{z-1} \ge p\\
\leftrightarrow &prob \ge p\\
\end{aligned} \nonumber
\end{equation}

In summary, $prob \ge p$ if and only if $s + q \le 1 $, conditioned on $z > 1$ and $p \in (0,1)$. It is easy to verify that the equality holds if and only if $s + q = 1 $, following the same derivation process.
\end{proof}

% \section{contour plot}
% \label{app:contour}

% In this section, we discuss 

% %In this section, we present the contour plot of $prob$ with respect to $q$ and $s$ under the conditions $z = 5$ and $p = 0.5$ in Fig.\ref{fig:contour}.

% \begin{figure}
%     \centering
%     \includegraphics[width=\linewidth]{fig/contour plot.pdf}    \caption{The contour plot of $prob$ with respect to $q$ and $s$ under the conditions $z = 5$ and $p = 0.5$.}
%     \label{fig:contour}
% \end{figure}

\section{Simulation Experiment}
\label{app:simulation}

In this section, we present a simulation study to verify the theoretical results in Theorem~\ref{th:kappa}. More specifically, we choose the values of the parameters $p$,$q$,$s$, and use these parameters to control the behavior of the simulated Actor and Critic. An example of the simulation process is as follows:

1. Actor simulation: sample a random value $k$ following a uniform distribution U(0,1). If $k < p$, then the sample $S$ is regarded as a correct SQL, otherwise a wrong SQL. This simulates the behavior of an Actor with accuracy of $p$.

2. Critic simulation - Branch 1: if $S$ is a correct SQL, then sample a value $k$ from U(0,1). If $k < s$, then output False, otherwise output True. This simulates the behavior of a Critic with FPR=$s$. 

3. Critic simulation - Branch 2: if $S$ is an incorrect SQL, then sample a value $k$ U(0,1). If $k < q$, then output True, otherwise output False. This simulates the behavior of an Critic with FNR=$q$. 

4. If the Critic in the step 2 or 3 outputs True, then the iteration stops, otherwise return to (1). This process iterates until the number of iterations reaches the pre-defined $z=5$.

5. After the iteration stops, the SQL $S$ generated so far is output by the AC process. Note that whether $S$ is correct or not is known. 

6. Run this process for 1,000,000 times, and compute the proportion that $S$ is correct, which is the accuracy of the AC system. 

For each configuration setting $(p,q,s)$, we conducted the simulation 100 times, and used the average value as the simulation result. The results are shown in Table~\ref{tab:the and exp} in the form of (theoretical result, simulation result). It can be seen that the accuracy obtained from the simulation matches the expected accuracy $prob$ derived from Theorem~\ref{th:kappa}.

\end{document}